\documentclass[runningheads]{svmult}

\usepackage{makeidx}   
\usepackage{graphicx}  
\usepackage{subeqnar}  
\usepackage{multicol}  
\usepackage{physprbb}  
\makeindex             

\begin{document}
\title*{Spin Glasses:  Still Complex After All These Years?}
\toctitle{Spin Glasses and Complexity}
%
%
\titlerunning{Spin Glasses and Complexity}
%
\author{D.L.~Stein}

\authorrunning{D.L.~Stein}
%
%
\institute{Departments of Physics and Mathematics, University of Arizona,
Tucson, AZ 85721 USA}

\maketitle              

\begin{abstract}
Spin glasses are magnetic systems exhibiting both quenched disorder and
frustration, and have often been cited as examples of `complex systems.'
In this talk I review some of the basic notions of spin glass physics, and
discuss how some of our recent progress in understanding their properties
might lead to new viewpoints of how they manifest `complexity'.
\end{abstract}

\section{Introduction}
\label{sec:intro}

This talk will probably be a change of pace for most of you; at least the
topic is mostly orthogonal to those covered by the other talks.  In
particular, I'll be discussing a {\it classical\/} statistical mechanical
problem.  The origins of the interactions that define the spin glass are of
course quantum mechanical; and quantum phenomena in many spin glass systems
have become an active area of study over the past decade.  Nevertheless,
many of the important phenomena observed down to very low temperatures in a
wide variety of spin glasses can be explained using classical statistical
mechanics -- or, more truthfully, {\it could\/} be explained if we could
figure out how to treat the enormous complications arising from the
quenched randomness inherent in these systems.  Because of these
complications, the most basic questions remain open, and the spin glass has
often been touted as a model example of a complex system.

In the absence of a universally agreed definition of `complex system', it
is as difficult to argue with that claim as it is to justify it.  Maybe
spin glasses -- as well as other systems discussed at this meeting -- are
merely `complicated systems'.  What I'll try to do in this talk is to
convey some of the flavor of spin glass physics, and to show why its
understanding requires the introduction of some new concepts and tools into
statistical mechanics.  You can then judge for yourself whether the
classical spin glass fits your own understanding of `complexity'.

\medskip

The talk will be divided into four parts:

\medskip

\noindent $\bullet$ What is a spin glass?

\smallskip

\noindent $\bullet$ Why are they interesting to physicists?

\smallskip

\noindent $\bullet$ What is the current level of our understanding?

\smallskip

\noindent $\bullet$ What -- if anything -- do they have to do with complexity?

\section{Brief Review of Spin Glasses}
\label{sec:review}

Spin glasses are systems with localized electronic magnetic moments whose
interactions are characterized by {\it quenched randomness\/}: a given
pair of localized moments (`spins' for short) have a roughly equal {\it
a priori\/} probability of having a ferromagnetic or an antiferromagnetic
interaction.  The prototype material is a dilute magnetic alloy, with a
small amount of magnetic impurity randomly substituted into the lattice of
a nonmagnetic metallic host; for example, {\it Cu\/}Mn or {\it Au\/}Fe.
However, insulators such as Eu$_x$Sr$_{1-x}$S, with $x$ roughly between .1
and .5, also display spin glass behavior.  The underlying physics governing
the random interactions differs for different classes of materials; for the
dilute magneitc alloys, it arises from the conduction electron-mediated
RKKY interactions between the localized moments.

Early experiments by Cannella and Mydosh \cite{CM72} indicated that a phase
transition occurred in {\it Au\/}Fe alloys: the low-field ac magnetic
susceptibility exhibited a cusp at a frequency-dependent temperature $T_f$.
Similar behavior has since been seen in other spin glasses, and has become
a signature feature of spin glass behavior.  At the same time, specific
heat curves show no singularities, but instead a smoothly rounded maximum
at a temperature slightly above $T_f$ (for a review, see \cite{BY86}).
Whether there exists a true thermodynamic phase transition to a
low-temperature spin glass phase remains an open question.

Neutron magnetic scattering data and other probes of magnetic structure
indicate that at low temperatures, the spins are frozen -- at least on
experimental timescales -- in random orientations.  Hence the name {\it
spin glass\/}: the magnetic disorder is reminiscent of the translational
disorder in the atomic arrangement of an ordinary glass. 

\section{Why Should We Care?}
\label{sec:why}

So why should we care?

\smallskip

\noindent $\bullet$ {\bf Because it's there.}  From the perspective of
condensed matter physics, any new class of condensed matter systems is
worth understanding.  Of course, some classes of systems are more
interesting than others.  By more interesting, I mean that they may have
great importance for technological application, and/or they give rise to
powerful new ideas, and perhaps new physical or mathematical tools.  Often
these ideas and tools are applicable to different kinds of condensed matter
systems, and perhaps to problems outside of condensed matter physics
altogether.  A well-known example is the broken gauge symmetry of
superconductors providing a `mass' to the photon, which was influential
in the uncovering of the Higgs mechanism in particle physics.

Spin glasses are likely to belong to such a class of systems; indeed, as
we'll describe below, they have already proved a fertile ground for
uncovering new ideas and techniques with potentially wide applicability.
But returning to the problem of spin glasses proper: unlike ordinary
glasses, which must be cooled sufficiently rapidly to avoid the crystalline
phase, the spin glass has no competing ordered phase. So if a thermodynamic
phase transition does exist, then the low temperature phase would truly be
an equilibrium condensed disordered phase -- a new state of matter.

\smallskip

\noindent $\bullet$ {\bf Statistical Mechanics of Disordered Systems.}
Homogeneous systems, such as crystals, uniform ferromagnets, and
superfluids, display spatial symmetries that greatly simplify their
physical and mathematical analyses.  The absence of such symmetries
enormously complicates the understanding of disordered systems like spin
glasses.  This may lead to new types of broken symmetries, a breakdown of
the thermodynamic limit for certain quantities, the emergence of new
phenomena such as chaotic temperature dependence, the need for creation of
new thermodynamic tools, and other unanticipated features to be described
below.  While it may not be necessary to completely revamp statistical
mechanics in order to understand disordered systems, as has sometimes been
suggested, it is at least necessary to carefully rethink some deeply held
assumptions.

\smallskip

\noindent$\bullet$ {\bf Applications to Other Areas.} Concepts that arose
in the study of spin glasses have led to applications in areas as diverse
as computer science \cite{KGV83,MP85,FA86,MP86}, neural networks \cite{Hop82,AGS85},
prebiotic evolution \cite{PWA82,AS84,RAS87}, protein conformational
dynamics \cite{Ste85}, protein folding \cite{BW90}, and a variety of others.
We will not have time to discuss these applications here, but extensive
treatments can be found in \cite{MPV87,Ste92,Nish01}.

\section{Spin Glass Theory}
\label{sec:theory}

The modern theory of spin glasses began with the work of Edwards and
Anderson (EA) \cite{EA75}, who proposed that the essential physics of spin
glasses lay not in the details of their microscopic interactions but rather
in the {\it competition\/} between quenched ferromagnetic and
antiferromagnetic interactions.  It should therefore be sufficient to study
the Hamiltonian
\begin{equation}
\label{eq:EA}
{\cal H}_{\cal J}=-\sum_{<x,y>} J_{xy} \sigma_x\sigma_y -h\sum_x\sigma_x\ ,
\end{equation}
where $x$ is a site in a $d$-dimensional cubic lattice, $\sigma_x=\pm 1$ is
the Ising spin at site $x$, $h$ is an external magnetic field, and the
first sum is over nearest neighbor sites only.  To keep things simple, we
take $h=0$ and the spin couplings $J_{xy}$ to be independent Gaussian
random variables whose common distribution has mean zero and variance one.
With these simplifications, the EA Hamiltonian~(\ref{eq:EA}) has global
spin inversion symmetry.  We denote by ${\cal J}$ a particular realization
of the couplings, corresponding physically to a specific spin glass sample.

\subsection{Frustration}
\label{subsec:frustration}

\begin{figure}[t]
\begin{center}
\includegraphics[width=.3\textwidth]{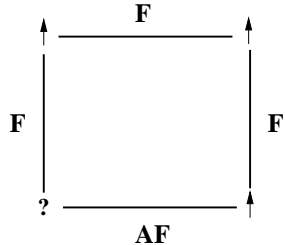}
\end{center}
\caption[]{A simple frustrated contour in a $2D$ lattice.  Bonds marked
``F'' correspond to ferromagnetic couplings ($J_{xy}>0$) and ``AF''
corresponds to an antiferromagnetic coupling ($J_{xy}<0$).  One possible
arrangement of spins at the corner sites is shown.}
\label{fig:frustration}
\end{figure}

The Hamiltonian~(\ref{eq:EA}) exhibits {\it frustration\/}: no spin
configuration can simultaneously satisfy all couplings.  If a closed
circuit ${\cal C}$ in the edge lattice satisfies the property 
\begin{equation}
\label{eq:frustration}
\prod_{<x,y>\in{\cal C}}J_{xy}<0\, .
\end{equation}
then the spins along it cannot all be simultaneously satisfied \cite{T77}
(Fig.~\ref{fig:frustration}).

Anderson \cite{And78} suggested a different formulation, namely that
frustration manifests itself as free energy fluctuations scaling as the
{\it square root\/} of the surface area of a typical sample.  Either way,
the spin glass is characterized by both {\it quenched disorder\/} and {\it
frustration\/}.  Their joint presence indicates the possibility that spin
glasses might possess multiple pure thermodynamic states unrelated by any
simple symmetry transformation.  We will return to this question later.

\subsection{Mean Field Theory}
\label{subsec:mft}

Within months of appearance of the EA model, an infinite-ranged version was
proposed by Sherrington and Kirkpatrick (SK) \cite{SK75}.  For a system of
$N$ Ising spins, and in zero external field, their Hamiltonian is

\begin{equation}
\label{eq:SK}
{\cal H}_{{\cal J},N}=-{1\over\sqrt{N}}\sum_{1\le i<j\le N}J_{ij} \sigma_i\sigma_j 
\end{equation}
where the independent, identically distributed couplings $J_{ij}$ are again
chosen from a Gaussian distribution with zero mean and variance one; the
$1/\sqrt{N}$ rescaling ensures a sensible thermodynamic limit for free
energy per spin and other thermodynamic quantities.

SK showed that their model had an equilibrium phase transition at $T_c=1$.
While the static susceptibility had a cusp there, so did the specific heat.
This wasn't necessarily surprising given that infinite-ranged models aren't
expected to correctly describe the behavior of low-dimensional systems at
the critical point.  More troubling was SK's observation that the
low-temperature phase had an instability: in particular, the entropy became
negative at very low temperature.

A mean field theory, employing the Onsager reaction field term, was
proposed two years later by Thouless, Anderson, and Palmer \cite{TAP77}.
Their approach indicated that there might be many low-temperature
solutions, possibly corresponding to different spin glass `phases'.  (As a
point of nomenclature, one should probably reserve use of the term `mean
field theory' for the TAP model.  Nevertheless, to save space and time, I
will follow general practice and use the term to refer to the
infinite-ranged SK model also.)  Other important early papers include the
work of deAlmeida and Thouless \cite{AT78}, who considered the stability of
the SK solution in the $h$-$T$ plane, and the dynamical work of Sompolinsky
and Zippelius \cite{SZ81,Somp81,SZ82}.

We will not have time to discuss these papers here, and will focus instead
on what is believed today to be the correct solution for the
low-temperature phase of the SK model.  This solution, due to Parisi
\cite{P79}, employed a novel {\it ansatz\/} and required several more years
before a physical interpretation could be worked out
\cite{P83,MPSTV84a,MPSTV84b}.  The picture that finally arose was that of a
system with an extraordinary new kind of symmetry breaking, known today as
`replica symmetry breaking', or RSB, after the mathematical procedures used
to derive it.  The essential idea is that the low-temperature phase
consists not of a single spin-reversed pair of states, but rather of
``infinitely many pure thermodynamic states'' \cite{P83}, not related by
any simple symmetry transformations.  In the next section, we describe the
qualitative features of the Parisi solution in greater detail.

\subsection{Broken Replica Symmetry}
\label{subsec:rsb}

It had been pointed out by EA that a correct description of the spin glass
phase needs to reflect the lack of orientational spin order with the spin
`frozenness', or long-range order in time.  Denoting by $\Lambda_L$ a cube
of side $L$ centered at the origin, and $\langle\cdot\rangle$ a thermal
average, the magnetization per spin in a pure phase
\begin{equation}
\label{eq:mag}
M=\lim_{L\to\infty}{1\over|\Lambda_L|}\sum_{x\in\Lambda_L}\langle\sigma_x\rangle
\end{equation}
should vanish (for a.e.~${\cal J}$), while
\begin{equation}
\label{eq:qea}
q_{EA}=\lim_{L\to\infty}{1\over|\Lambda_L|}\sum_{x\in\Lambda_L}\langle\sigma_x\rangle^2
\end{equation}
should not.

The quantity $q_{EA}$ measures the breaking of time-reversal symmetry, and
is now known as the `EA order parameter', but by itself is not sufficient
to describe the broken symmetry of the SK spin glass phase.  The correct
order parameter needs to describe the structure and relationships among the
infinitely many states present at low temperature.  (We ignore here the
problems inherent in defining `pure state' for the SK model; for more
discussion on this, see \cite{NS02,NSreview,NSinprep}.)  To do this, we
consider the {\it overlap\/} between two states $\alpha$ and $\beta$, at
fixed ${\cal J}$ and $T$:

\begin{equation}
\label{eq:qabSK}
q_{\alpha\beta}\approx {1\over
N}\sum_{i=1}^N\langle\sigma_i\rangle_\alpha\langle\sigma_i\rangle_\beta\, ,
\end{equation}
where $\langle\cdot\rangle_\alpha$ is a thermal average in pure state
$\alpha$.

Given the infinity of states, quantities referring to individual pure
states are of little use, even if such things could be defined.  What is
really of interest is the {\it distribution\/} of overlaps.  Consider
choosing two pure states randomly and independently from the Gibbs
distribution at fixed $N$; this will be a mixture over many pure states
$\alpha$, with varying weights $W_{\cal J}^{\alpha}$ (dependence on $N$ and
$T$ is suppressed for ease of notation).  Let $P_{\cal J}(q)dq$ be the
probability that the overlap of the two states lies between $q$ and $q+dq$.
$P_{\cal J}(q)$ is commonly referred to as the {\it Parisi overlap
distribution\/}.  It is equal to 
\begin{equation}
\label{eq:ovdistSK}
P_{\cal J}(q)=\sum_\alpha\sum_\beta W_{\cal J}^{\alpha}W_{\cal
J}^{\beta}\delta(q-q_{\alpha\beta})\, .
\end{equation}

If there is a single pure state, such as the paramagnet at $T>T_c$, then
$P_{\cal J}(q)$ is simply a $\delta$-function at $q=0$.  For ferromagnets
with free or periodic boundary conditions, there are only two pure states,
namely the uniform positive and negative magnetization states, each
appearing in any finite-volume Gibbs state with weight $1/2$.  The overlap
distribution function is now a pair of $\delta$-functions, each with weight
$1/2$, located at $\pm M^2(T)$.

\begin{figure}[b]
\begin{center}
\includegraphics[width=.3\textwidth]{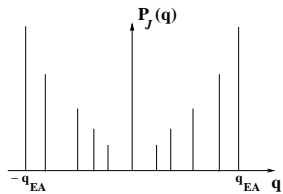}
\end{center}
\caption[]{Sketch of the overlap distribution function $P_{\cal J}(q)$ for
the SK model below $T_c$.}
\label{fig:SKoverlap}
\end{figure}

What about in the SK model?  According to the Parisi solution, for fixed
${\cal J}$ and (large) $N$, it has the form qualitatively sketched in
Fig.~\ref{fig:SKoverlap}.  The nontrivial nature of the overlap structure
reflects the presence of many states (although only a handful have weights
of $O(1)$) that are not related to each other by a simple symmetry
transformation.

Even more interesting is the {\it non-self-averaging\/} of the overlap
distribution function.  Suppose a new coupling realization ${\cal J'}$ is
considered.  Now, for any large $N$, the overlaps (except for the two at
$\pm q_{EA}$, which are present for almost every ${\cal J}$) will appear at
different values of $q$, and the set of corresponding weights will also
differ.  This is true no matter how large $N$ becomes.  

\begin{figure}[t]
\begin{center}
\includegraphics[width=.3\textwidth]{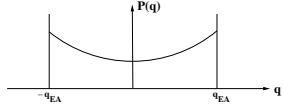}
\end{center}
\caption[]{Sketch of the averaged overlap distribution function $P(q)$ for
the SK model below $T_c$.  The spikes at $\pm q_{EA}$ are
$\delta$-functions.}
\label{fig:Pq}
\end{figure}

Let $P_N(q)=\overline{P_{{\cal J},N}(q)}$, where $\overline{[\cdot]}$ denotes
an average over coupling realizations, and $P(q)=\lim_{N\to\infty}P_N(q)$.
The resulting distribution $P(q)$ will be supported on all values of $q$ in
the interval $[-q_{EA},q_{EA}]$; a sketch is shown in Fig.~\ref{fig:Pq}.

Together $P_{{\cal J}}(q)$ and $P(q)$ can be thought of as describing the
broken symmetry of the SK model below $T_c$.  There is another famous
feature, namely the {\it ultrametricity\/} of the overlaps.  Instead of
considering pairs of states, one now considers triples.  The Parisi
solution predicts that, for almost every fixed ${\cal J}$, the distances
$d_{\alpha\beta}=q_{EA}-q_{\alpha\beta}$ among the three pairs will form
the sides of an equilateral or an acute isosceles triangle.  These strong
correlations among the overlaps correspond to a tree-like, or hierarchical,
structure of the states.

So the infinite-ranged spin glass has a strikingly novel type of broken
symmetry, much different from anything observed in homogeneous systems.
Because mean-field theory has usually provided a reliable description of
the {\it low-temperature\/} properties of finite-dimensional models, it was
perhaps natural to expect that the RSB mean-field picture should similarly
describe the nature of ordering in the EA and other short-ranged spin glass
models.  We now turn to these and related issues.

\subsection{Some open questions}
\label{subsec:questions}

We saw from the preceding discussion that, already in mean field theory,
spin glasses possess unusual and exotic properties.  But what has been
presented so far is only a piece of the story.  Laboratory spin glasses
display an array of puzzling thermodynamic and dynamical behavior.  We
confine ourselves here to a discussion of the nature of broken symmetry and
ordering of the low-temperature spin glass phase -- if there is one.  At
the same time, we emphasize that this represents only a part of the overall
picture.

In this section we list some open -- and fundamental -- questions regarding
the nature of broken symmetry in the short-ranged spin glass.

\smallskip

$\bullet$  Is there a phase transition to a low-temperature spin glass
phase?

\smallskip

We have already seen that the answer is `yes' for the infinite-ranged
model.  For the EA and other short-ranged models, the answer is not
definitively known.  There is some analytical \cite{FS90,TH96} and
numerical \cite{BY86,O85,OM85,KY96} work that supports existence of a phase
transition in three -- and even more likely, four -- dimensional Ising spin
glasses.  But the issue remains unsettled \cite{MPR94}.  It is usually
assumed (though it doesn't necessarily follow) that a low-temperature phase
will be non-unique and display broken spin-flip symmetry -- that is, a
nonzero $q_{EA}$.

If there is no broken-spin-flip symmetric spin glass phase in any finite
dimension, then the study of spin glasses becomes one of dynamics.  In what
follows we will simply assume that such a phase does exist, above some
lower critical dimension.  Then an important question is:

\smallskip 

$\bullet$ Are there infinitely many pure state pairs below $T_c$ in the EA
spin glass?

\smallskip

If RSB correctly describes the low-$T$ phase, then the answer is yes.
However, a competing picture \cite{Mac84,BM85,BM87,FH86,HF87a,FH87b,FH88}
that arose in the early to mid-80's, based on domain-wall renormalization
group ideas, leads to a very different picture of the low-$T$ phase.  This
approach, known as droplet/scaling, leads to the conclusion that there is
only a {\it single pair\/} of spin-flip-reversed pure states at low
temperature in {\it any\/} finite dimension.

Which of these alternatives is the case is not known in any dimension
greater than one (where of course there is only a single pair).  In two
dimension, it is believed that $T_c=0$, so the issue becomes one of the
number of {\it ground\/} state pairs only.  Recent numerical experiments
\cite{Mid99,PY99a,Hart99} support the possibility of only a single pair of
ground states.  Recent rigorous work \cite{NS00,NS01a} also supports the
notion that only a single pair of ground states occurs in two dimensions.
In three dimensions numerical simulations give conflicting results
\cite{PY99b,MP00}.

But let's suppose that there are infinitely many ground and/or pure states
above some lower critical dimension.  Even if that's the case, it does not
follow that the mean-field picture of replica symmetry breaking holds in
finite dimensions, because that picture requires a very intricate pattern
of relationships among all of the states.  So we come to our third question:

\smallskip

$\bullet$ If there do exist infinitely many equilibrium states in some
finite dimensions, is their organization similar to that of the Parisi
solution of the SK model?

\smallskip

Here the question {\it has\/} largely been answered, due to a series of
rigorous and heuristic results.

\subsection{Constraints on the Ordering of Short-Ranged Spin Glasses}
\label{subsec:constraints}

In a series of papers \cite{NS02,NS96a,NS96b,NSBerlin,NS97,NS98}, Newman
and Stein have shown that the complex structure of replica symmetry
breaking cannot be supported by finite-dimensional spin glasses in any
dimension and at any temperature.  The arguments will not be presented
here, but instead we will focus on alternative scenarios that remain
viable.

\smallskip

One possibility is that $T_c=0$ in any finite dimension, and all of the
interesting features of spin glasses arise from dynamical processes.  It
could also happen that $T_c>0$, but there remains a unique
(non-paramagnetic) Gibbs state below $T_c$.  While either of these
possibilities may end up being the case, it is the feeling of most workers
in the field that there is a low-temperature phase with broken spin-flip
symmetry.

If this is so, then \cite{NS96a,NS96b,NSBerlin,NS97,NS98} leave open two
main possibilities.  (I emphasize here that these are not the only
remaining {\it logical\/} possibilities, but rather the most plausible
ones.)  The first is a two-state picture like droplet/scaling, described in
Sect.~\ref{subsec:questions}.  What about many-state pictures?  There is
one that would be consistent with the rigorous results of
\cite{NS96a,NS96b,NS98}, called the {\it chaotic pairs\/} picture
\cite{NS96b,NS97,NS98,NS92}.

Consider a cube $\Lambda_L$ with $L$ large, and with periodic boundary
conditions.  In the chaotic pairs picture, the resulting finite-volume
Gibbs state for any $0<T<T_c$ would consist of a single pair of
spin-flip-related pure states, just as in the droplet/scaling picture.  But
in the latter picture the {\it same\/} pair of pure states appears in every
large volume, while in chaotic pairs the pure state pair appearing in
$\Lambda_L$ will vary chaotically with $L$.

So chaotic pairs resembles the droplet/scaling picture in any single finite
volume, but differs over a collection of volumes and hence has a different
thermodynamic structure.  It is a many-state picture, but unlike the
mean-field picture, it has a {\it trivial\/} overlap function.  In any
finite volume, $P_{\cal J}(q)$ would be a pair of $\delta$-functions at
$\pm q_{EA}$, just as in the droplet/scaling picture.  If one instead
constructed the overlap function of the infinite-volume pure states, then
it would most likely be a single $\delta$-function at the origin, for
almost every ${\cal J}$. So, even though there are infinitely many states
in this picture, there is no nontrivial replica symmetry breaking, no
non-self-averaging, and no ultrametricity.

\section{Are Spin Glasses Complex Systems?}
\label{sec:complex}

The preceding overview enables us to return to the issue posed in the title
of the talk.  As I promised early on, I won't attempt to define
`complexity' or `complex system', but will instead review some of the
salient properties of spin glasses and leave it to you to decide whether
they fit into your conception of a complex system.  Whatever your answer,
perhaps a more important characterization is whether you find them to be
interesting and possibly relevant to problems that you work on.  (And if
the answer to all these is no, I hope that the talk at least was a pleasant
diversion!)

I should begin by emphasizing that I did not cover, or even mention, many
of the features of spin glasses that years ago helped to earn them the
title of `complex system'.  These include some of the following: their
property of displaying many metastable states, that is, states stable to
flips of finite numbers of spins; their possessing a `rugged energy
landscape' (more or less equivalent to the preceding property, but
sometimes also used to denote the presence of many pure or ground states);
and their anomalous dynamical behaviors --- slow relaxation,
irreversibility, memory effects, hysteresis, and aging.  I did briefly
touch on their connections to problems in computer science, biology, and
other areas, and of course that's important also.

But I'd like to emphasize here some of their more newly discovered
properties that perhaps have not received as much attention.  In what
follows, I list several features --- some recently uncovered, some not ---
in which spin glasses display unexpected behavior (`unexpected' meaning
`not familiar from our experience with the statistical mechanics of
homogeneous systems').  A similar discussion appears in \cite{NSreview},
and much of it represents joint work with Chuck Newman.

The appearance of broken replica symmetry in the infinite-ranged spin glass
alone might merit the `complexity' label, especially given its hierarchical
structure in state space.  A hierarchical organization of components has
often been used to explain how a complex organization can be built out of
simple components \cite{Simon73}.  However, it was noted above that this
type of broken symmetry is absent in short-ranged spin glass models.

But this in itself is an interesting piece of news, because mean field
theory has almost always served as an invaluable guide to the
low-temperature behavior of statistical mechanical systems, in particular
the nature of the order parameter and broken symmetry.  The failure of mean
field theory to provide a correct description of the low-temperature phase
in {\it any\/} finite dimension indicates that the $d\to\infty$ limit of
the EA model is singular.  This possibility was broached by Fisher and Huse
\cite{FH87b}, and our work confirms their conjecture.

The failure of mean-field theory to describe the broken symmetry of
short-ranged models is perhaps just as interesting --- and, if you're so
inclined, just as good a candidate for the label `complex' --- as the
exotic features of the Parisi solution.  But {\it why\/} does mean field
theory fail for realistic spin glasses? This is discussed at length in
\cite{NSreview,NSinprep}, and I'll refer the interested reader to those
papers.  For those who are moderately interested, but not enough to begin
reading yet another paper on this subject, I'll just note here that one
important reason lies in the combination of the physical couplings scaling
to zero in the SK model, along with their statistical independence.  This
combination ensures that something like the following must happen: suppose
that some fixed $N_1$ corresponds to a particular `pure state' structure
(roughly speaking).  As $N$ continues to increase, it will eventually reach
an $N_2\gg N_1$ in which the earlier pure state structure, corresponding to
$N_1$, is completely washed out.  Any fixed, finite set of couplings will
eventually have no effect on the spin glass state for $N$ large enough.  In
contrast, short-ranged models do not share this peculiar feature, or at
least it should be considerably weaker there.  For these models, the
couplings outside of a particular fixed, finite region can act at most on
its boundaries.

A second important feature is the possible nonexistence of a
`straightforward' thermodynamic limit for Gibbs states.  By
`straightforward' I mean that a sequence of finite-volume Gibbs states,
generated along an infinite sequence of volumes with boundary conditions
chosen independently of the couplings, may not yield a limiting
thermodynamic state.  This should occur whenever there are many pure state
pairs \cite{NS92}, although I haven't had time to discuss it here.  But
it's a reflection of the lack of any spatial symmetries that allow one to
choose simple boundary conditions, like free or periodic, or an external
symmetry-breaking field, that can lead to the existence of such a limit.
(Two caveats: first, one can always choose subsequences that {\it do\/}
lead to limiting thermodynamic states, but they would presumably have to be
chosen in a coupling-dependent way, and at present there are no known ways
of doing so.  Second, this discussion is confined to Gibbs states, or
equivalently correlation functions, which depend sensitively on the local
details of the coupling realization.  Global quantities, such as the free
energy per spin, {\it do\/} have a limit for a given coupling-independent
sequence of boundary conditions, for a.e.~${\cal J}$.)

This has an interesting consequence, related to our usual expectations for
the behavior of large condensed matter systems.  Statistical mechanical
calculations typically rely on the assumption that the thermodynamic limit
reveals the bulk properties of macroscopic systems.  And in fact it almost
certainly does, for disordered systems as well as homogeneous.  But if, in
the former case, there exist many competing pure states, then the
connection between large (but finite) and infinite systems may be far more
subtle than in homogeneous systems, where a straightforward extrapolation
is generally sufficient.  In order to connect the thermodynamic behavior of
disordered systems with that of large but finite volumes, we have found it
extremely useful to utilize a new thermodynamic tool which we call the {\it
metastate\/} \cite{NS96b,NSBerlin,NS97,NS98,AW90},

Finally, it is possible that spin glasses display {\it chaotic temperature
dependence\/} \cite{BM87,FH88}, in which correlation functions on
sufficiently large lengthscales change in an unpredictable (though
deterministic) fashion under infinitesimal temperature changes.

So the study of spin glasses has provided us with a host of new phenomena
and tools: analytical tools such as replica symmetry breaking for
infinite-ranged systems, and metastates for both short-ranged and
infinite-ranged systems; numerical tools such as simulated annealing; and
others.  The infinite-ranged spin glass displays a beautiful new type of
order, and the short-ranged spin glass seems to break many of the old rules
and assumptions.  Perhaps most surprisingly, analytical -- and even
rigorous --- progress has been achievable.  As this progress continues,
there is little doubt that more surprises remain in store.

\noindent{\it Acknowledgments.} I wish to thank Chuck Newman for a long and
enjoyable collaboration on these questions, and for helpful comments on the
manuscript.  This work was supported in part by National Science Foundation
Grant DMS-01-02541.

\end{document}